\documentclass[12pt, a4paper]{article}  
\usepackage{amsmath}
\usepackage[ansinew]{inputenc}
\usepackage{typearea,epsfig,rotating,times}
\typearea{12}
\input{epsf.tex}
\allowdisplaybreaks
\begin{document}
\newcommand{\bi}[1]{\bibitem{#1}}
\def\be#1\ee{\begin{equation}#1\end{equation}}
\def\bea{\begin{eqnarray}}
\def\eea{\end{eqnarray}}
\newcommand\vr{\boldsymbol{r}}
\newcommand\vc{\boldsymbol{c}}
\newcommand\vh{\boldsymbol{a}} 
\newcommand\vv{\boldsymbol{v}}
\newcommand\valpha{\boldsymbol{\alpha}}

\null\bigskip

{\large \centerline{Physics Notes\footnote{Edited by C.E.~Baum,
      Air Force Research Laboratory, Kirtland Airforce Base, 
New Mexico, USA }}\medskip

\centerline{Note 13}
\medskip
\centerline{28 June 2005}}
\vspace*{1.5cm}

{\LARGE
\centerline{Axiomatics of classical electrodynamics} 
\centerline{and its relation to gauge field theory}}

\vspace*{0.8cm}

\large \centerline{Frank Gronwald${}^1$, Friedrich W.\ Hehl${}^{2,3}$,
  and J\"urgen Nitsch${}^1$}

\vspace{0.8cm}

{\normalsize
\it 
\centerline{${}^1$ Otto-von-Guericke-University of Magdeburg}
\centerline{Institute for Fundamental Electrical Engineering and EMC}
\centerline{Post Box 4120, 39016 Magdeburg, Germany}
\bigskip
\centerline{${}^2$ University of Cologne}
\centerline{Institute for Theoretical Physics}
\centerline{50923 K\"oln, Germany}
\bigskip
\centerline{${}^3$ University of Missouri-Columbia}
\centerline{Department of Physics and Astronomy}
\centerline{Columbia, MO 65211, USA}}
\bigskip

\normalsize\centerline{{\it file essential14.tex}}

\bigskip
\bigskip

\normalsize
\centerline{\bf Abstract}
\noindent We give a concise axiomatic introduction into the fundamental
structure of classical electrodynamics: It is based on electric
charge conservation, the Lorentz force, magnetic flux conservation,
and the existence of local and linear constitutive relations. The {\it
  inhomogeneous\/} Maxwell equations, expressed in terms of $D^i$ and
$H_i$ , turn out to be a consequence of electric charge conservation,
whereas the {\it homogeneous\/} Maxwell equations, expressed in terms
of $E_i$ and $B^i$, are derived from magnetic flux conservation and
special relativity theory. The excitations $D^i$ and $H_i$ , by means
of constitutive relations, are linked to the field strengths $E_i$ and
$B^i$. Eventually, we point out how this axiomatic approach is related
to the framework of gauge field theory.

\vfill
\noindent
{\scriptsize E-Mail: Frank.Gronwald@et.uni-magdeburg.de,
  hehl@thp.uni-koeln.de, Juergen.Nitsch@et.uni-magdeburg.de}

\newpage

\normalsize

\section*{Table of contents}
\begin{description}

\item[1] Introduction

\item[2] Essential classical electrodynamics based on four axioms
\begin{description}
\item[2.1] Electric charge conservation (axiom 1) and the inhomogeneous
  Maxwell equations
\item[2.2] Lorentz force (axiom 2) and merging of electric and
  magnetic field strengths
\item[2.3] Magnetic flux conservation (axiom 3) and the homogeneous
  Maxwell equations
\item[2.4] Constitutive relations (axiom 4) and the properties of
  spacetime
\end{description}

\item[3]  On the relation between the axiomatics and the gauge approach
\begin{description}
\item[3.1] Noether theorem and electric charge conservation
\item[3.2] Minimal coupling and the Lorentz force
\item[3.3] Bianchi identity and magnetic flux conservation
\item[3.4] Gauge approach and constitutive relations
\end{description}

\item[4] Conclusion

\item \hspace{-6pt}Acknowledgments

\item \hspace{-6pt}Appendix: Mathematical background
\begin{description}

\item[A.1] Integration
\begin{description}
\item[A.1.1] Integration over a curve and covariant vectors as line
  integrands
\item[A.1.2] Integration over a surface and contravariant vector
  densities as surface integrands
\item[A.1.3] Integration over a volume and scalar densities as volume
  integrands
\end{description}

\item[A.2] Poincar\'e lemma

\item[A.3] Stokes theorem

\end{description}
\item \hspace{-6pt}References

\end{description}

\newpage
\normalsize
\section{Introduction}\label{sec1}

In nature one has, up to now, identified four fundamental
interactions: Gravity, electromagnetism, weak interaction, and strong
interaction. Gravity and electromagnetism manifest themselves on a
macroscopic level. The weak and the strong interactions are
generically microscopic in nature and require a quantum field
theoretical description right from the beginning.

The four interactions can be modeled individually. Thereby it is
recognized that electromagnetism has the simplest structure amongst
these interactions.  This simplicity is reflected in the Maxwell
equations. They, together with a few additional assumptions, explain
the electromagnetic phenomena that we observe in nature or in
laboratories.

Without digressing to philosophy, one may wonder about the origin of
the Maxwell equations. Should we believe in them as such and just
study their consequences? Or should we rather derive them from some
deeper lying structures?  Certainly, there are already some answers
known to the last question. The Maxwell equations rely on conservation
laws and symmetry principles that are also known from elementary
particle physics, see \cite{chen84,ryde96}.  In the framework of
classical physics, authoritative accounts of electrodynamics are
provided by \cite{rohr65,schw98}, e.g.. In this paper we would like to
add some new insight into this subject.

We will provide a short layout of an axiomatic approach that allows to
identify the {\it basic ingredients\/} that are necessary for
formulating classical electrodynamics, see \cite{hehl03}. We believe
that this axiomatic approach is not only characterized by simplicity
and beauty, but is also of appreciable pedagogical value. The more
clearly a structure is presented, the easier it is to memorize.
Moreover, an understanding of how the fundamental electromagnetic
quantities $D^i,\,H_i,\,E_i,\,B^i$ are related to each other may
facilitate the formulation and solution of actual electromagnetic
problems.

As it is appropriate for an axiomatic approach, we will start from as
few prerequisites as possible. What we will need is some elementary
mathematical background that comprises differentiation and integration
in the framework of tensor analysis in three-dimensional space. In
particular, the concept of integration is necessary for introducing
electromagnetic objects as integrands in a natural way. To this end,
we will use a tensor notation in which the components of mathematical
quantities are explicitly indicated by means of upper (contravariant)
or lower (covariant) indices \cite{scho89}.  The advantage of this
notation is that it allows to represent geometric properties clearly.
In this way, the electromagnetic objects become more transparent and
can be discussed more easily. For the formalism of differential
forms, which we recommend and which provides similar conceptual
advantages, we refer to \cite{lind04,hehl03}.
 
We have compiled some mathematical material in the Appendix. Those who
don't feel comfortable with some of the notation, may first want to
have a look into the Appendix. Let us introduce the following
conventions:
\begin{itemize}
\item Partial derivatives with respect to a spatial coordinate $x^i$
  (with $i,j,\dots=1,2,3$) or with resepct to time $t$ are abbreviated
  according to \be \frac{\partial}{\partial x^i} \;\longrightarrow\;
  \partial_i\,, \qquad\qquad \frac{\partial}{\partial {t}}
  \;\longrightarrow\; \partial_{{\rm t}} \,.  \ee
\item We use the ``summation convention''. It states that a summation
  sign can be omitted if the same index occurs both in a lower and an
  upper position. That is, we have, for example, the correspondence
  \be \sum_{i=1}^3 \alpha_i\,\beta^i \;\longleftrightarrow\;
  \alpha_i\, \beta^i \,.  \ee
\item We define the Levi-Civita symbols $\epsilon_{ijk}$ and
  $\epsilon^{ijk}$.  They are antisymmetric with respect to all of
  their indices. Therefore, they vanish if two of their indices are
  equal. Their remaining components assume the values $+1$ or $-1$,
  depending on whether $ijk$ is an even or an odd permutation of
  $123$: \be \epsilon_{ijk} = \epsilon^{ijk}=
\begin{cases}
\hspace{8pt}1\,, & \text{for $ijk$ = 123, 312, 231}, \\
-1\,, & \text{for $ijk$ = 213, 321, 132}. \end{cases}
\ee
\end{itemize}
With these conventions we obtain for the {\sl gradient\/} of a
function $f$ the expression $\partial_i f$. The {\sl curl\/} of a
(covariant) vector $v_i$ is written according to
$\epsilon^{ijk}\partial_j v_k$ and the {\sl divergence\/} of a
(contravariant) vector (density) $w^i$ is given by $\partial_i w^i$.

Now we are prepared to move on to the Maxwell theory.

\section{Essential classical electrodynamics based on four axioms}
\label{sec:axiomatic}

In the next four subsections, we will base classical electrodynamics
on electric charge conservation (axiom~1), the Lorentz force
(axiom~2), magnetic flux conservation (axiom~3), and the existence of
constitutive relations (axiom~4). This represents the core of
classical electrodynamics: It results in the Maxwell equations
together with the constitutive relations and the Lorentz force law.

In order to complete electrodynamics, one can require two more axioms,
which we only mention shortly (see \cite{hehl03} for a detailed
discussion). One can specify the energy-momentum distribution of the
electromagentic field (axiom~5) by means of its so-called
energy-momentum tensor. This tensor yields the energy density $(D^iE_i
+H_iB^i)/2$ and the energy flux density $\epsilon^{ijk}E_jH_k$ (the
Poynting vector), inter alia.  Moreover, if one treats electromagnetic
problems of materials in macrophysics, one needs a further axiom by
means of which the total electric charge (and the current) is split
(axiom~6) in a bound or material charge (and current), which is also
conserved, and in a free or external charge (and current). This
completes classical electrodynamics.

\subsection{Electric charge conservation (axiom 1) and the 
inhomogeneous Maxwell equations}

In classical electrodynamics, the electric charge is characterized by
its density~$\rho$. From a geometric point of view, the charge density
$\rho$ constitutes an integrand of a volume integral.  This geometric
identification is natural since, by definition, integration of $\rho$
over a three-dimensional volume $V$ yields the total charge $Q$
enclosed in this volume \be Q:=\int_V \rho \, dv \,.  \ee We note that,
in the SI-system, electric charge is measured in units of ``ampere
times second'' or coulomb, $[Q]=\text{As}=\text{C}$.  Therefore the
SI-unit of charge density $\rho$ is
$[\rho]=\text{As}/\text{m}^3=\text{C}/ \text{m}^3$.

It is instructive to invoke at this point the Poincar\'e lemma.  There
are different explicit versions of this lemma. We use the form
\eqref{poin3} that is displayed in Appendix~\ref{sec:background}.
Then (if space fulfills suitable topological conditions) we can write
the charge density $\rho$ as the divergence of an integrand $D^i$ of a
surface integral. Thus, \be \boxed{ \partial_i D^i = \rho\,
\label{inhom1}
} \qquad \qquad( {\rm div}\,{\cal D}=\rho) \,.  \ee This result
already constitutes one inhomogeneous Maxwell equation, the
Coulomb-Gauss law. In parenthesis we put the symbolic form of this
equation.

Electric charges often move. We represent this motion by a material
velocity field $u^i$, that is, we assign locally a velocity to each
portion of charge in space. The product of electric charge density
$\rho$ and material velocity $u^i$ defines the electric current
density $J^i$, \be J^i= \rho u^i\,.
\label{currentdef}
\ee 
Geometrically, the electric current density constitutes an integrand
of surface integrals since integration of $J^i$ over a two-dimensional
surface $S$ yields the total electric current $I$ that crosses this
surface, \be I=\int_S J^i \, da_i \,.  \ee We have, in SI-units,
$[I]=\text{A}$ and $[J^i]=\text{A}/\text{m}^2$.

We now turn to electric charge conservation, the first axiom of our 
axiomatic approach. To this end we have to determine how individual
packets of charge change in time as they move with velocity $u^i$ through 
space. A convenient
way to describe this change is provided by the material derivative 
$D/Dt$ which also is often called convective derivative~\cite{roth01}. It
allows to calculate the change of a physical quantity as it appears
to an observer or a probe that follows this quantity. 
Then electric charge conservation can be expressed as
\be
{
\frac{DQ}{Dt} = 0\,,
}
\label{axiom1}
\ee
where the material derivative is taken with respect to the velocity 
field $u^i$. It can be rewritten in the following way \cite{roth01}, 
\begin{align}
\frac{DQ}{Dt} & = \frac{D}{Dt}\int_{V(t)}\rho\,dV\nonumber \\ 
&=\int_{V(t)}\frac{\partial \rho}{\partial t}
\, dV + \oint_{\partial V(t)} \rho u^i\, da_i\nonumber \\
&=\int_{V(t)} \left(\frac{\partial \rho}{\partial t} + 
\partial_i(\rho  u^i)\right)\, dV\,.
\label{intcont}
\end{align}
Here we used in the last line the Stokes theorem in the form of 
\eqref{stokes1}.  The volume $V(t)$ that is integrated over depends in
general on time since it moves together with the electric charge
that it contains. By means of \eqref{currentdef}, \eqref{axiom1}, and
\eqref{intcont} we obtain the axiom of electric charge conservation in
the local form as continuity equation, \be \partial_{\rm t} \rho +
\partial_i J^i = 0 \,.
\label{cont1}
\ee
 
Now we use the inhomogeneous Maxwell equation \eqref{inhom1} in order 
to replace 
within the continuity equation \eqref{cont1} the charge density 
by the divergence of $D^i$. This yields
\be
\partial_i\Bigl(\partial_{\rm t} D^i + J^i\Bigr) = 0\,.
\label{cont2}
\ee
Again we invoke the Poincar\'e lemma, now in the form \eqref{poin2}, 
and write the sum $\partial_{\rm t} D^i + J^i$ as
the curl of the integrand of a line integral which we denote by $H_i$.
This yields
\be
\boxed{
\epsilon^{ijk}\partial_j H_k - \partial_{\rm t} D^i  =  J^i \,}
\label{inhom2}
\qquad\quad ({\rm curl}\,{H}-\dot{\cal D}={\cal J})\,.  \ee Equation
\eqref{inhom2} constitutes the remaining inhomogeneous Maxwell
equation, the Amp\`ere-Maxwell law, which, in this way, is derived
{}from the axiom of charge conservation.  The fields $D^i$ and $H_i$ are
called electric excitation (historically: electric displacement)
and magnetic excitation (historically: magnetic field),
respectively. From \eqref{inhom1} and \eqref{inhom2} it follows that
their SI-units are $[D^i]=\text{As}/\text{m}^2$ and
$[H_i]=\text{A}/\text{m}$.

Some remarks are appropriate now: We first note that we obtain the
excitations $D^i$ and $H_i$ from the Poincar\'e lemma and charge
conservation, respectively, without introducing the concept of force.
This is in contrast to other approaches that rely on the Coulomb and
the Lorentz force laws \cite{elli92}. Furthermore, since electric
charge conservation is valid not only on macroscopic scales but also
in micropysics, the inhomogeneous Maxwell equations \eqref{inhom1} and
\eqref{inhom2} are microphysical equations as long as the source terms
$\rho$ and $J^i$ are microscopically formulated as well. The same is
valid for the excitations $D^i$ and $H_i$. They are microphysical
quantities --- in contrast to what is often stated in textbooks, see
\cite{jack98}, for example.  We finally remark that the inhomogeneous
Maxwell equations \eqref{inhom1} and \eqref{inhom2} can be
straightforwardly put into a relativistic\-ally invariant form. This
is not self-evident but suggested by electric charge conservation in
the form of the continuity equation \eqref{cont1} since this
fundamental equation can also be shown to be relativistically
invariant.

\subsection{Lorentz force (axiom 2) and merging of electric and
  magnetic field strengths}

During the discovery of the electromagnetic field, the concept of
force has played a major role. Electric and magnetic forces are
directly accessible to experimental observation. Experimental evidence
shows that, in general, an electric charge is subject to a force if an
electromagnetic field acts on it. For a point charge $q$ at position
${x_{\rm q}}^i$, we have $\rho(x^i) = q \delta(x^i-{x_{\rm q}}^i)$. If
it has the velocity $u^i$, we postulate the Lorentz force \be \boxed{
  F_i=q(E_i+\epsilon_{ijk}u^jB^k) \,
\label{lorentzforce}
} \ee as second axiom. It introduces the electric field strength $E_i$
and the magnetic field strength $B^i$. The Lorentz force already
yields a prescription of how to measure $E_i$ and $B^i$ by means of
the force that is experienced by an infinitesimally small test charge
$q$ which is either at rest or moving with velocity $u^i$. Turning to
the dimensions, we introduce voltage as ``work per charge''. In SI, it
is measured in volt (V). Then $[F_i]$=VC/m and, according to
\eqref{lorentzforce}, $[E^i]=\text{V}/\text{m}$ and
$[B_i]=\text{Vs}/\text{m}^2 =\text{Wb}/\text{m}^2=\text{T}$, with Wb as
abbreviation for weber and T for tesla.

\vspace{0.5cm}
\begin{figure}[htb]
\begin{center}
\includegraphics[width=0.9\linewidth]{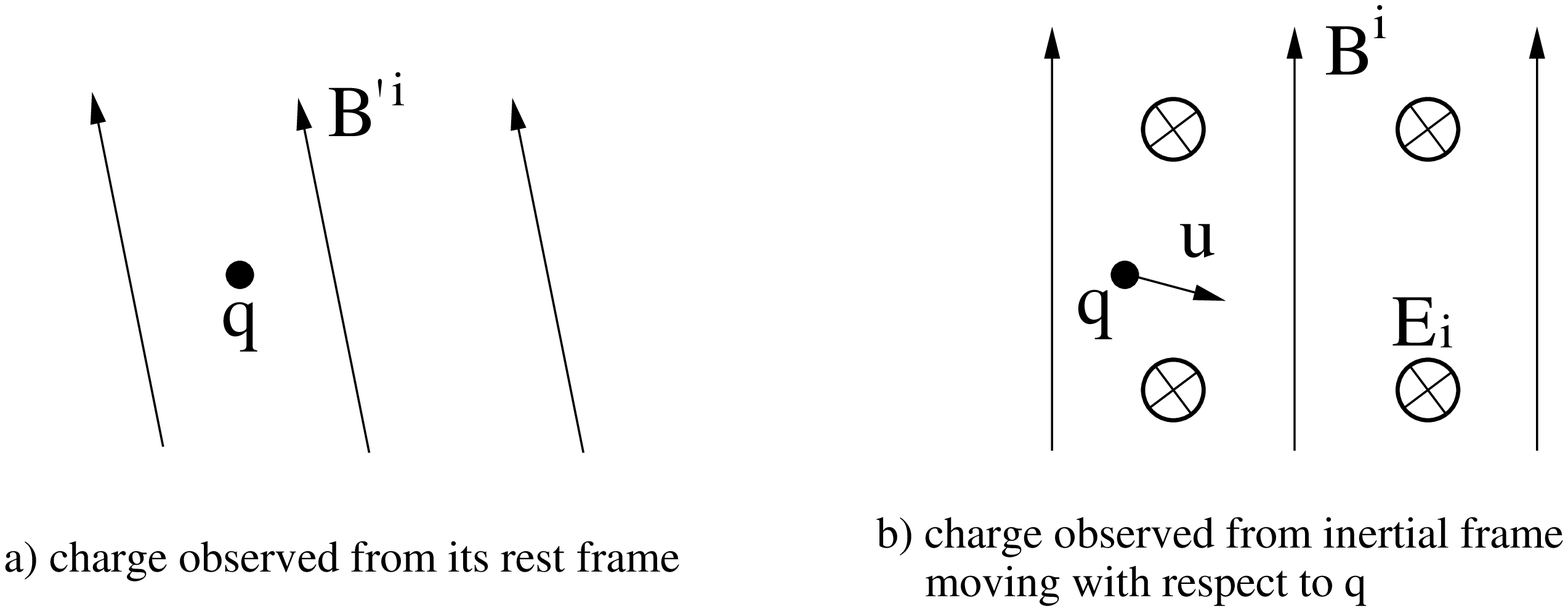}
\caption{A charge that is, in some inertial frame, at rest and is 
  immersed in a purely magnetic field experiences no Lorentz force,
  see Fig.1a.  The fact that there is no Lorentz force should be
  independent of the choice of the inertial system that is used to
  observe the charge.  Therefore, a compensating electric field
  accompanies the magnetic field if viewed from an inertial laboratory
  system which is in relative motion to the charge, see Fig.1b.}
\label{relative1}
\end{center}
\end{figure}

{}From the axiom of the Lorentz force \eqref{lorentzforce}, we can draw
the conclusion that the electric and the magnetic field strengths are
not independent of each other. The corresponding argument is based
on the special relativity principle: According to the special
relativity principle, the laws of physics are independent of the
choice of an inertial system \cite{elli92}.  Different inertial
systems move with constant velocities $v^i$ relative to each other.
The outcome of a physical experiment, as expressed by an empirical
law, has to be independent of the inertial system where the experiment
takes place.

Let us suppose a point charge $q$ with a certain mass moves with
velocity $u^i$ in an electromagnetic field $E_i$ and $B^i$. The
velocity and the electromagnetic field are measured in an inertial
laboratory frame.  The point charge can also be observed from its
instantaneous inertial {\it rest frame.} If we denote quantities that
are measured with respect to this rest frame by a prime, i.e., by
${u'}^i$, ${E'}_i$, and ${B'}^i$, then we have ${u'}^i=0$. In the
absence of an electric field in the laboratory system, i.e., if
additionally ${E'}_i=0$, the charge experiences no Lorentz force and
therefore no acceleration, \be {F'}_i = q({E'}_i +
\epsilon_{ijk}{u'}^j{B'}^k ) = 0\,.  \ee

\vspace{0.5cm}
\begin{figure}[htb]
\begin{center}
\includegraphics[width=0.4\linewidth]{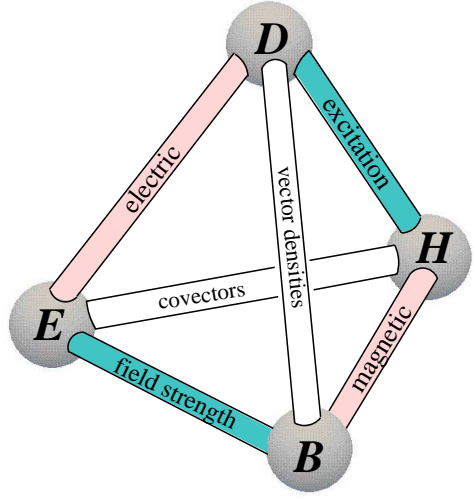}
\caption{The tetrahedron of the electromagnetic field. The
electric and the magnetic excitations $D^i,H_i$ and the electric and
the magnetic field strengths $E_i,B^i$ build up 4-dimensional
quantities in spacetime. These four fields describe the
electromagnetic field completely. Of electric nature are ${D^i}$ and
$E_i$, of magnetic nature $H_i$ and $B^i$.  The electric and the
magnetic excitations $D^i,H_i$ are extensities, also called quantities
(how much?), the electric and the magnetic field strengths $E_i,B^i$
are intensities, also called forces (how strong?).}
\label{tetrahedron4}
\end{center}
\end{figure}

The fact that the charge experiences no acceleration is also true in
the laboratory frame. This is a consequence of the special relativity
principle or, more precisely, of the fact that the square of the
acceleration can be shown to form a relativistic invariant.
Consequently, \be F_i = q(E_i + \epsilon_{ijk}u^j B^k ) = 0\,.  \ee
Thus, in the laboratory frame, electric and magnetic field are related
by \be E_i=-\epsilon_{ijk}u^jB^k \,.
\label{compensate}
\ee 
This situation is depicted in Fig.\ref{relative1}.
Accordingly, we find that electric and magnetic field strength
cannot be viewed as independent quantities. They are connected to each
other by transformations between different inertial systems.

Let us pause for a moment and summarize: So far we have introduced the
four electromagnetic field quantities $D^i,H_i$ and $E_i,B^i$. These
four quantities are interrelated by physical and mathematical
properties.  This is illustrated in Fig.\ref{tetrahedron4} by the
``tetrahedron of the electromagnetic field''.

\subsection{Magnetic flux conservation (axiom 3) and the
homogeneous Maxwell equations}

We digress for a moment and turn to hydrodynamics. Helmholtz was one
of the first who studied rotational or ``vortex'' motion in
hydrodynamics, see \cite{lamb36}. He derived theorems for vortex
lines. An important consequence of his work was the conclusion that
vortex lines are conserved. They may move or change orientation but
they are never spontaneously created nor annihilated. The vortex lines
that pierce through a two-dimensional surface can be integrated over
and yield a scalar quantity that is called circulation.  The
circulation in a perfect fluid, which satisfies certain conditions, is
constant provided the loop enclosing the surface moves with the fluid
\cite{lamb36}.

There are certainly fundamental differences between electromagnetism
and hydrodynamics. But some suggestive analogies exist. A vortex line
in hydrodynamics seems analogous to a magnetic flux line.  The
magnetic flux $\Phi$ is determined from magnetic flux lines,
represented by the magnetic field strength $B^i$, that pierce through
a two-dimensional surface $S$, \be \Phi := \int_S B^i\, da_i\,.
\label{flux}
\ee As the circulation in a perfect fluid is conserved, we can guess
that, in a similar way, the magnetic flux may be conserved. Of course,
the consequences of such an axiom have to be borne out by experiment. 

At first sight, one may find vortex lines of a fluid easier to
visualize than magnetic flux lines. However, on a microscopic level,
magnetic flux can occur in quanta. The corresponding magnetic flux
unit is called flux quantum or fluxon and it carries $\Phi_0 =
h/(2e)\approx 2,07\cdot 10^{-15}$~Wb, with $h$ as Planck constant and
$e$ as elementary charge. Single quantized magnetic flux lines have
been observed in the interior of type II superconductors if exposed to
a sufficiently strong magnetic field, see \cite{hehl03}, p.131. They
even can be counted. The corresponding experiments provide good
evidence that magnetic flux is a conserved quantity.

But how can we formulate magnetic flux conservation mathematically? It
is at this point instructive to reconsider the notion of the electric
charge \be Q=\int_V \rho \, dv\, \ee together with its corresponding
conservation law \be \partial_{\rm t} Q + \int_{\partial V} J^i \,
da_i= 0 \,.
\label{global1}
\ee The rate of change of the electric charge within a specified
volume $V$ is balanced by the out- or inflowing charge across the
surface $\partial V$. This charge transport is described by the
electric charge {\it current\/} $J^i$ that is integrated over the
enveloping surface $\partial V$. By means of the Stokes theorem in the
form \eqref{stokes1}, equation \eqref{global1} yields the local
continuity equation \be \partial_{\rm t} \rho + \partial_i J^i = 0 \,.
\ee

Let us follow the same pattern in formulating magnetic flux
conservation: Starting with the definition \eqref{flux} of the
magnetic flux, the corresponding conservation law, in analogy to
\eqref{global1}, reads \be \partial_{\rm t} \Phi + \int_{\partial S}
J_i^\Phi \, dc^i = 0\,,
\label{global2}
\ee where we introduced the magnetic flux {\it current\/} $J_i^\Phi$.
Geometrically, this is a covariant vector that is integrated along a
line $\partial S$, that is, along the curve bordering the
2-dimensional surface $S$.  The conservation law \eqref{global2} tells
us that the rate of change of the magnetic flux within a specified area
$S$ is balanced by the magnetic flux current $J_i^\Phi$ that is
integrated along the boundary $\partial S$. Then the Stokes
theorem in the form \eqref{stokes2} yields 
the local continuity equation \be \partial_{\rm t} B^i +
\epsilon^{ijk}\partial_j J_k^\Phi = 0\,.
\label{local2}
\ee

One interesting consequence is the following: The divergence of
\eqref{local2} reads \be \partial_i(\partial_{\rm t} B^i)=0 \qquad
\Longrightarrow \qquad \partial_i B^i = \rho_{\rm mag}\,,\quad
\partial_{\rm t} \rho_{\rm mag}=0\,.
\label{consequence}
\ee Thus, we find a time-{\em in\/}dependent term $\rho_{\rm mag}$,
which acquires tentatively the meaning of a magnetic charge density.
Let us choose a specific reference system in which $\rho_{\rm mag}$ is
constant in time, i.e., $\partial_{\rm t} \rho_{\rm mag}=0$. Now we go
over to an arbitrary reference system with time coordinate $t'$ and
spatial coordinates $x^{i'}$. Clearly, in general $\partial_{\rm t'}
\rho_{\rm mag}\ne 0$.  The only way to evade a contradiction to
(\ref{consequence}) is to require $\rho_{\rm mag}=0$, that is, the
magnetic field strength $B^i$ has no sources, its divergence vanishes:
\be \boxed{ \partial_i B^i = 0\,
\label{homo1}
}\qquad\quad({\rm div}\,{\cal B}=0)\,.  \ee This is recognized as one
of the homogeneous Maxwell equations. Note that our derivation of
\eqref{homo1} was done under the assumption of magnetic flux
conservation \eqref{global2}. Under this condition we find $\rho_{\rm
  mag}=0$.

In order to understand better the magnetic flux current, we note that
$J_i^\Phi$, as a covariant vector, has the same geometric properties
as the electric field strength $E_i$. Additionally, $J_i^\Phi$ and
$E_i$ share the same physical dimension voltage/length, i.e., in SI,
$\text{V}/\text{m}$. Accordingly, it is plausible to identify both
quantities, \be J_i^\Phi \equiv E_i \,.
\label{identification}
\ee That also the sign chosen is the appropriate one (consistent with
the Lenz rule) was discussed in \cite{ItinHehl}. Then the local
continuity equation \eqref{local2} assumes the form \be \boxed{
  \partial_{\rm t} B^i + \epsilon^{ijk} \partial_j E_k = 0 \,
\label{homo2}
}\qquad\quad (\dot{\cal B}+{\rm curl}\,E=0)\,.  
\ee 
This equation reflects magnetic flux conservation, the third axiom of
our axiomatic approach. It also constitutes the remaining homogeneous Maxwell
equation, that is, Faraday's induction law.

At this point one might wonder to what extend the identification
\eqref{identification} is mandatory. It turns out that it is special
relativity that dictates this identification. We illustrate this
circumstance as follows: In the rest frame of a magnetic flux line
${B'}^i$ the magnetic flux current vanishes, ${J'}^\Phi_i = 0$. The
rest frame is also defined via the Lorentz force: In the absence of an
electric field, ${E'}_i=0$, a test charge $q$ is not accelerated by
${B'}^i$. Then a Lorentz transformation, together with \eqref{homo1},
yields an equation that relates $B^i$ and $J^\Phi_i$ in a laboratory
frame according to \be J^\Phi_i=-\epsilon_{ijk}u^jB^k \,.
\label{compensate2}
\ee A comparison with \eqref{compensate}, which was obtained by an
analogous transformation of a magnetic flux line from its rest frame
to a laboratory frame, shows that the identification
\eqref{identification} needs to be valid, indeed.  However, one should
be aware that our simple argument requires ${E'}_i=0$ in the rest
frame of the considered magnetic flux line.

\vspace{-8pt}

\subsection{Constitutive relations (axiom 4) and the properties
  of spacetime}

So far we have introduced $4\times 3=12$ unknown electromagnetic field
components $D^i, H_i$, $E_i$, and $B^i$. These components have to
fulfill the Maxwell equations \eqref{inhom1}, \eqref{inhom2},
\eqref{homo1}, and \eqref{homo2}, which represent $1+3+1+3=8$ partial
differential equations. In fact, among the Maxwell equations, only
\eqref{inhom2} and \eqref{homo2} contain time derivatives and are
dynamical. The remaining equations, \eqref{inhom1} and \eqref{homo1},
are so-called ``constraints''.  They are, by virtue of the dynamical
Maxwell equations, fulfilled at all times if fulfilled at one time.
It follows that they don't contain information on the time evolution
of the electromagnetic field.  Therefore, we arrive at only 6
dynamical equations for 12 unknown field components.  To make the
Maxwell equations a determined set of partial differential equations,
we still have to introduce additionally the so-called ``constitutive
relations'' between the excitations $D^i$, $H_i$ and the field
strengths $E_i$, $B^i$.

The simplest case to begin with is to find constitutive relations for
the case of electromagnetic fields in vacuum. There are guiding
principles that limit their structure. We demand that constitutive
relations in vacuum are invariant under translation and rotation,
furthermore they should be local and linear, i.e., they should connect
fields at the same position and at the same time.  Finally, in vacuum the
constitutive relations should not mix electric and magnetic
properties.  These features characterize the vacuum and not the
electromagnetic field itself. We will not be able to prove them but
postulate them as fourth axiom.

If we want to relate the field strengths and the excitations we have
to remind ourselves that $E_i$, $H_i$ are natural integrands of {\it
  line\/} integrals and $D^i$, $B^i$ are natural integrands of {\it
  surface\/} integrals. Therefore, $E_i$, $H_i$ transform under a
change of coordinates as covariant vectors while $D^i$, $B^i$
transform as contravariant vector densities.  To compensate these
differences we will have to introduce a symmetric metric field
$g_{ij}=g_{ji}$.  The metric tensor determines spatial distances and
introduces the notion of orthogonality. The determinant of the metric
is denoted by $g$. It follows that $\sqrt{g}g^{ij}$ transforms like a
density and maps a covariant vector into a contravariant vector
density.  We then take as fourth axiom the constitutive equations for
vacuum, 
\begin{equation} \label{const1} 
  \hspace{-22pt} \boxed{
D^i=\varepsilon_0 \,\sqrt{g}\, g^{ij}\, E_j \,,}
\end{equation}
\begin{equation}\label{const2} 
 \boxed{ 
 H_i=(\mu_0\,\sqrt{g})^{-1}g_{ij} \, B^j
    \,.}
\end{equation} 
In flat spacetime and in cartesian coordinates, we have $g=1$,
$g^{ii}=1$, and $g^{ij}=0$ for $i\neq j$. We recognize the familiar
vaccum relations between field strengths and excitations.  The
electric constant $\varepsilon_0$ and the magnetic constant $\mu_0$
characterize the vacuum. They acquire the SI-units
$[\varepsilon_0]=\text{As}/\text{Vm}$ and
$[\mu_0]=\text{Vs}/\text{Am}$.

What seems to be conceptually important about the constitutive
equations \eqref{const1}, \eqref{const2} is that they not only provide
relations between the excitations $D^i$, $H_i$ and the field strengths
$E_i$, $B^i$, but also connect the electromagnetic field to the
structure of spacetime, which here is represented by the metric tensor
$g_{ij}$. The formulation of the first three axioms that were
presented in the previous sections does not require information on
this metric structure.  The connection between the electromagnetic
field and spacetime, as expressed by the constitutive equations,
indicates that physical fields and spacetime are not independent of
each other.  The constitutive equations might suggest the point of
view that the structure of spacetime determines the structure of the
electromagnetic field.  However, one should be aware that the opposite
conclusion has a better truth value: It can be shown that the
propagation properties of the electromagnetic field determine the
metric structure of spacetime \cite{hehl03,laemm}.
 
Constitutive equations in matter usually assume a more complicated
form than \eqref{const1}, \eqref{const2}. In this case it would be
appropriate to derive the constitutive equations, after an averaging
procedure, from a microscopic model of matter. Such procedures are the
subject of solid state or plasma physics, for example. A discussion of
these subjects is out of the scope of this paper but, without going
into details, we quote the constitutive relations of a general linear
{\it magnetoelectric\/} medium:
\begin{eqnarray}\label{explicit'}
  {D}^i\!&=\!&\left(\,
    \boldsymbol{\varepsilon^{{ij}}}\hspace{4pt} - \,
    \epsilon^{ijk}\,n_k \right)E_j\,+\left(\hspace{9pt}
    \boldsymbol{\gamma^i{}_j} + \tilde{s}_j{}^i\right) {B}^j +
  (\boldsymbol{\alpha}-s)\,B^i \,,\\ {H}_i\!  &=\!  &\left(
    \boldsymbol{ \mu_{ij}^{-1}} - {\epsilon}_{ijk}\,m^k \right) {B}^j
  +\left(- \boldsymbol{ \gamma^j{}_i} + \tilde{s}_i{}^j \right)E_j -
  (\boldsymbol{\alpha}+s)\,E_i\,.\label{explicit''}
\end{eqnarray}
This formulation is due to Hehl \& Obukhov
\cite{hehl03,PostCon,measure}, an equivalent formulation of a
``bianisotropic medium'' --- this is the same as what we call general
linear medium --- was given by Lindell \& Olyslager
\cite{exotic,lind04}.  Both matrices $\varepsilon^{ij}$ and
$\mu_{ij}^{-1}$ are symmetric and possess 6 independent components
each, $\varepsilon^{ij}$ is called {\it permittivity\/} tensor and
$\mu^{-1}_{ij}$ {\it impermeability\/} tensor (reciprocal permeability
tensor). The magnetoelectric cross-term $\gamma^i{}_j$, which is
tracefree, $\gamma^k{}_k=0$, has 8 independent components. It is
related to the Fresnel-Fizeau effects.

The 4-dimensional pseudo-scalar $\alpha$, we call it axion piece
\cite{hehl03}, represents one component. It corresponds to the perfect
electromagnetic conductor (PEMC) of Lindell \& Sihvola
\cite{LindSihv2004a}, a Tellegen type structure
\cite{Tellegen1948,Tellegen1956/7}.

Accordingly, these pieces altogether, which we printed in
(\ref{explicit'}) and (\ref{explicit''}) in boldface for better
visibility, add up to $6+6+8+1=20+1=21$ independent components.  The
situation with 20 components is described in Post \cite{Post} (he
reqiured $\alpha=0$ without a real proof), that with 21 components in
O'Dell \cite{O'Dell}.

We can have 15 more components related to dissipation, which can{\it not\/} be
derived from a Lagrangian, the so-called skewon piece (see
\cite{skewon} and the literature given), namely $3+3$ components of
$n_k$ and $m^k$ (electric and magnetic Faraday effects), 8 components
{}from the matrix $\tilde{s}_i{}^j$ (optical activity), which is
traceless $\tilde{s}_k{}^k=0$, and 1 component {}from the
3-dimensional scalar $s$ (spatially isotropic optical activity). This
scalar was introduced by Nieves \& Pal \cite{NP94}.  It has also been
discussed in electromagnetic materials as chiral parameter, see
Lindell et al.\ \cite{Lindell1994}. Note that $s$, in contrast to the
4-dimensional scalar $\alpha$, is only a 3D scalar. We end then up
with the {\it general linear medium\/} with $20+1+15=36$ components.

With the introduction of constitutive equations the axiomatic approach
to classical electrodynamics is completed. We will see in the next
Section \ref{sec:relation} how this approach relates to the framework
of gauge theory.

\section{On the relation between the axiomatics and the gauge approach}
\label{sec:relation}

Modern descriptions of the fundamental interactions heavily rely on
symmetry principles. In particular, this is true for the
electromagnetic interaction which can be formulated as a gauge field
theory that is based on a corresponding gauge symmetry. In a recent
article this approach towards electromagnetism has been explained in
some detail \cite{gron01}.  The main steps were the following:
\begin{itemize}
\item Accept the fact that physical matter fields (which represent
  electrons, for example) are described microscopically by complex
  wave functions.
\item Recognize that the absolute phase of these wave functions has no
  physical relevance. This arbitrariness of the absolute phase
  constitutes a one-dimensional rotational type symmetry $U(1)$ (the
  circle group) that is the gauge symmetry of electromagnetism.
\item To derive observable physical quantities from the wave functions
  requires to define derivatives of wave functions in a way that is
  invariant under the gauge symmetry.  The construction of such
  ``gauge covariant'' derivatives requires the introduction of gauge
  potentials.  One gauge potential, the scalar potential $\phi$,
  defines a gauge covariant derivative $D^\phi_{\rm t}$ with respect
  to time, while another gauge potential, the vector potential $A_i$,
  defines gauge covariant derivatives $D^A_i$ with respect to the
  three independent directions of space.
\item Finally, the gauge potentials $\phi$ and $A_i$ describe an
  electrodynamically non-trivial situation, if their corresponding
  electric and magnetic field strengths
\begin{align}
E_i &= -\partial_i \phi - \partial_{\rm t} A_i \,, \label{connect1}\\
B^i &= \hspace{10pt}\varepsilon^{ijk}\partial_j A_k \,, \label{connect2}
\end{align}
are non-vanishing.
\end{itemize}
In the following we want to comment on the interrelation between the
previously presented axiomatic approach and the gauge approach. It is
interesting to see how the axioms find their proper place within the
gauge approach.

\subsection{Noether theorem and electric charge conservation}

In field theory there is a famous result which connects symmetries 
of laws of nature to conserved quantities. This is the Noether theorem
which has been proven to be useful in both classical and quantum contexts.
It is, in particular, discussed in books on classical electrodynamics,
see \cite{rohr65,schw98}, for example.

Laws of nature, like in electrodynamics, e.g., can often (but not
always) be characterized concisely by a Lagrangian density ${\cal L}=
{\cal L}(\Psi, \partial_i \Psi, \partial_{\rm t} \Psi)$ which, in the
standard case, is a function of the fields $\Psi$ of the theory and
their first derivatives. Integration of the Lagrangian density ${\cal
  L}$ over space yields the Lagrangian $L$, \be L = \int {\cal
  L}(\Psi, \partial_i \Psi, \partial_{\rm t} \Psi)\, dV\,, \ee and
further integration over time yields the action $S$, \be S = \int L \,
dt\,.  \ee There are guiding principles that tell us how to obtain an
appropriate Lagrangian density for a given theory. Once we have an
appropriate Lagrangian density, we can derive conveniently the
properties of the fields $\Psi$. For example, the equations of motion
which determine the dynamics of $\Psi$ follow from extremization of
the action $S$ with respect to variations of $\Psi$, \be \delta_\Psi S
= 0 \qquad \Longrightarrow \qquad \text{equations of motion for
  $\Psi$}\,.
\label{motion}
\ee 

Now we turn to the Noether theorem which connects the symmetry of a
Lagrangian density ${\cal L}(\Psi, \partial_i \Psi, \partial_{\rm t}
\Psi)$ to conserved quantities. Suppose that ${\cal L}$ is invariant
under time translations $\delta_{\rm t}$. In daily life this
assumption makes sense since we do not expect that the laws of nature
change in time. Then the Noether theorem implies a local conservation
law which expresses the conservation of energy.  Similarly, invariance
under translations $\delta_{x^i}$ in space implies conservation of
momentum, while invariance under rotations $\delta_{\omega_i{}^j}$
yields the conservation of angular momentum,
\begin{align}
\delta_{\rm t} {\cal L} = 0 \qquad &\Longrightarrow \qquad 
\text{conservation of energy} \,,\\
\delta_{x^i} {\cal L} = 0 \qquad &\Longrightarrow \qquad 
\text{conservation of momentum} \,,\\
\delta_{\omega_i{}^j} {\cal L} = 0 \qquad 
&\Longrightarrow \qquad \text{conservation of angular momentum}\,.
\end{align}
These symmetries of spacetime are called external symmetries.
But the Noether theorem also works for other types of symmetries,
so-called internal ones --- especially gauge symmetries. In this case,
gauge invariance of the Lagrangian implies a conserved current with an
associated charge. That is, if we denote a gauge transformation by
$\delta_\epsilon$ we conclude \be \delta_{\epsilon} {\cal L} = 0
\qquad \Longrightarrow \qquad \text{charge conservation} \,.  \ee If
we apply this conclusion to electrodynamics, we have to specify the
Lagrangian density to be the one of matter fields that represent
electrically charged particles. Then invariance of this Lagrangian
density under the gauge symmetry of electrodynamics yields the
conservation of electric charge. Thus, if we accept the validity of
the Lagrangian formalism, then we can arrive at electric charge
conservation from gauge invariance via the Noether theorem.

\subsection{Minimal coupling and the Lorentz force}

We already have mentioned that, according to \eqref{motion}, we can
derive the equations of motion of a physical theory from a Lagrangian
density and its associated action. We can use this scheme to derive
the equations of motion of electrically charged particles.  In this
case, the corresponding Lagrangian density (that of the electrically
charged particles) has to be gauge invariant.

If electrically charged particles are represented by their
wave functions, the corresponding Lagrangian density will contain derivatives
with respect to time and space. It follows that the Lagrangian density 
will be gauge invariant if we pass from partial derivatives to
gauge covariant derivatives according to 
\begin{align}
  \partial_{\rm t} \qquad &\longrightarrow \qquad D^\phi_{\rm t} :=
  \partial_{\rm t}+ \frac{q}{\hbar}\phi\,,
\label{coupling1}\\
\partial_i \qquad &\longrightarrow \qquad D^A_i := 
\partial_i - \frac{q}{\hbar}A_i \,, 
\label{coupling2}
\end{align}
with $q$ the electric charge of a particle, $\hbar = h/(2\pi)$ with
$h$ as the Planck constant and $\phi$, $A_i$ as
electromagnetic potentials \cite{gron01}.  
This enforcement of gauge invariance has
a classical analogue. If electrically charged particles are
represented by point particles, rather than by wave functions, we have
to replace within the Lagrangian density the energy $E$ and the
momentum $p_i$ of each particle according to \cite{schw98}
\begin{align}
E \qquad &\longrightarrow \qquad E + q\phi \,,
\label{coupling3} \\
p_i \qquad &\longrightarrow \qquad p_i - q A_i \,.
\label{coupling4}
\end{align}
The substitutions \eqref{coupling1}, \eqref{coupling2} or
\eqref{coupling3}, \eqref{coupling4} constitute the simplest way to
ensure gauge invariance of the Lagrangian density of electrically
charged particles. They constitute what commonly is called ``minimal
coupling''. Due to minimal coupling, we relate electrically charged
particles and the electromagnetic field in a natural way that is
dictated by the requirement of gauge invariance.

Having ensured gauge invariance of the action $S$, we can derive 
equations of motion by extremization, compare \eqref{motion}. 
It then turns out that these equations of motion contain 
the Lorentz force law \eqref{lorentzforce}. Therefore the Lorentz force 
is a consequence of the minimal coupling procedure which couples 
electrically charged particles to the electromagnetic potentials and
makes the Lagrangian gauge invariant.

\subsection{Bianchi identity and magnetic flux conservation}

The electromagnetic gauge potentials $\phi$ and $A_i$ are often
introduced as mathematical tools to facilitate the integration of the
Maxwell equations.  Indeed, if we put the relations \eqref{connect1}
and \eqref{connect2} into the homogeneous Maxwell equations
\eqref{homo1} and \eqref{homo2}, we recognize that the homogeneous
Maxwell equations are fulfilled automatically. They become mere
mathematical identities. This is an interesting observation since
within the gauge approach the gauge potentials are fundamental
physical quantities and are not only the outcome of a mathematical
trick.  Thus we can state that the mathematical structure of the gauge
potentials already implies the homogeneous Maxwell equations and, in
turn, magnetic flux conservation. In this light, magnetic flux
conservation, within the gauge approach, appears as the consequence of
a geometric identity. This is in contrast to electric charge
conservation that can be viewed as the consequence of gauge
invariance, i.e., as the consequence of a physical symmetry.

The mathematical identity that is reflected in the homogeneous Maxwell
equations is a special case of a ``Bianchi identity''. Bianchi identities are 
the result of differentiating a potential twice. For example, in
electrostatics the electric field strength $E_i$ can be derived from a
scalar potential $\phi$ according to
\be
E_i = \partial_i \phi\,.
\ee
Differentiation reveals that the curl of $E_i$ vanishes,
\be
\epsilon^{ijk}\partial_j E_k = \epsilon^{ijk}\partial_j\partial_k \phi = 0\,,
\ee
which is due to the antisymmetry of $\epsilon^{ijk}$. Again, this equation 
is a mathematical identity, a simple example of a Bianchi identity.

\subsection{Gauge approach and constitutive relations}

The gauge approach towards electrodynamics deals with the properties
of gauge fields, which represent the electromagnetic field, and with
matter fields.  It does not reflect properties of spacetime. In
contrast to this, the constitutive equations do reflect properties of
spacetime, as can be already seen from the constitutive equations of
vacuum that involve the metric $g_{ij}$, compare \eqref{const1} and
\eqref{const2}. Thus, also in the gauge approach the constitutive
equations have to be postulated as an axiom in some way.  One should
note that, according to \eqref{connect1}, \eqref{connect2}, the gauge
potentials are directly related to the field strengths $E_i$ and
$B^i$. The excitations $D^i$ and $H_i$ are part of the inhomogeneous
Maxwell equations which, within the gauge approach, are derived as
equations of motion from an action principle, compare~\eqref{motion}.
Since the action itself involves the gauge potentials, one might
wonder how it is possible to obtain equations of motion for the
excitations rather than for the field strengths. The answer is that
during the construction of the action from the gauge potentials the
constitutive equations are already used, at least implicitly.

\section{Conclusion}
We have presented an axiomatic approach to classical electrodynamics
in which the Maxwell equations are derived from the conservation of
electric {\it charge\/} and magnetic {\it flux\/}. In the context of
the derivation of the inhomogenous Maxwell equations, one introduces
the electric and the magnetic excitation $D^i$ and $H_i$,
respectively.  The explicit calculation is rather simple because the
continuity equation for electric charge is already relativistically
invariant such that for the derivation of the inhomogeneous Maxwell
equations no additional ingredients from special relativity are
necessary. The situation is slightly more complicated for the
derivation of the homogeneous Maxwell equations from magnetic flux
conservation since it is not immediately clear of how to formulate
magnetic flux conservation in a relativistic invariant way. 
It should be mentioned that if the complete framework of relativity were 
available, the derivation of the axiomatic approach could be done with 
considerable more ease and elegance \cite{hehl03}.

Finally, we would like to comment on a question that sometimes leads
to controversial discussions, as summarized in \cite{roth01}, for
example. This is the question of how the quantities $E_i$, $D^i$,
$B^i$, and $H_i$ should be grouped in pairs, i.e., the question of
``which quantities belong together?''. Some people like to form the
pairs $(E_i,B^i)$, $(D^i, H_i)$, while others prefer to build
$(E_i,H_i)$ , $(D^i, B^i)$.  Already from a dimensional point of view,
the answer to this question is obvious. Both, $E_i$ and $B^i$ are {\it
  voltage}-related quantities, that is, related to the notions of
force and work: In SI, we have $[E_i]=\text{V/m}$,
$[B^i]=\text{T=Vs/m}^2$, or $[B^i]= [E_i]$/velocity.  Consequently,
they belong together.  Analogously, $D^i$ and $H_i$ are {\it
  current}-related quantities: $[D^i]=\text{C/m}^2=\text{As/m}^2$,
$[H_i]=\text{A/m}$, or $[D^i]= [H_i]$/velocity. Thermodynamically
speaking, $(E_i,B^i)$ are intensities (answer to the question: how
strong?) and  $(D^i, H_i)$ extensities (how much?)

These conclusions are made irrefutible by relativity theory.
Classical electrodynamics is a relativistic invariant theory and the
implications of relativity have been proven to be correct on macro-
and microscopic scales over and over again.  And relativity tells us
that the electromagnetic field strengths $E_i$, $B^i$ are inseparably
intertwined by relativistic transformations, and the same is true for
the electromagnetic excitations $D^i$, $H_i$. In the spacetime of
relativity theory, the pair $(E_i,B^i)$ forms one single quantity, the
tensor of electromagnetic field strength, while the pair $(D^i, H_i)$
forms another single quantity, the tensor of electromagnetic
excitations. If compared to these facts, arguments in favor of the
pairs $(E_i,H_i)$, namely that both are covectors, and $(D^i, B^i)$,
both are vector densities (see the tetrahedron in Fig.2), turn out to
be of secondary nature. Accordingly, there is no danger that the
couples $(E_i,B^i)$ and $(D^i, H_i)$ ever get divorced.

\subsection*{Acknowledgments}

We are grateful to Yakov Itin (Jerusalem), Ismo Lindell (Helsinki),
Yuri Obukhov (Cologne/ Moscow), and to G\"unter Wollenberg (Magdeburg)
for many interesting and helpful discussions.

\appendix

\section{Mathematical Background}
\label{sec:background}

Within a theoretical formulation physical quantities are modeled as
mathematical objects. The understanding and application of appropriate
mathematics yields, in turn, the properties of physical quantities. In
the development of the axiomatic approach, we made repeated use of
integration, of the Poincar\'e lemma, and of the Stokes theorem. It is with
these mathematical concepts that it is straightforward to derive the 
basics of electromagnetism from a small number of axioms.

\subsection{Integration}

Integration is an operation that yields coordinate independent values. It
requires an integration measure, the dimension of which depends on the
type of region that is integrated over. We want to integrate over
one-dimensional curves, two-dimensional surfaces, or three-dimensional
volumes that are embedded in three-dimensional space. Therefore, we
have to define line-, surface-, and volume-elements as
integration measures. Then we can think of suitable objects as
integrands that can be integrated over to yield coordinate independent
physical quantities.

\subsubsection{Integration over a curve and covariant vectors
as line integrands}

\begin{figure}[htb]
\begin{center}
\includegraphics[width=0.6\linewidth]{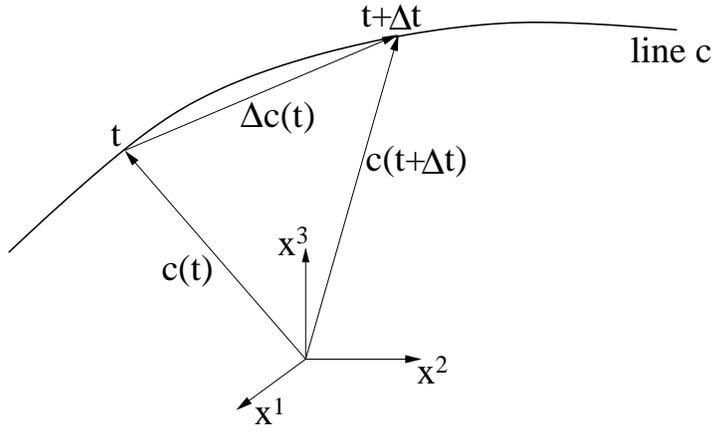}
\caption{Parametrization of a curve $\vc(t)$. The difference vector 
  $\Delta \vc(t)$ between $\vc(t+\Delta t)$ and $\vc(t)$ yields, in
  the limit $\Delta t\rightarrow 0$, the line element $d\vc(t)$.}
\label{curve1}
\end{center}
\end{figure}

We consider a one-dimensional curve $\vc=\vc(t)$ in three-dimensional
space. In a specific coordinate system $x^i$, with indices $i=1,2,3$,
a parametrization of $\vc$ is given by the vector \be
\vc(t)=\bigl(c^1(t), c^2(t), c^3(t)\bigr)\,.  \ee The functions
$c^i(t)$ define the shape of the curve. For small changes of the
parameter $t$, with $t\rightarrow t+\Delta t$, the difference vector
between $\vc(t+\Delta t)$ and $\vc(t)$ is given by \be \Delta \vc(t) =
\left(\frac{\Delta c^1}{\Delta t}, \frac{\Delta c^2}{\Delta t},
  \frac{\Delta c^3}{\Delta t}\right) \,\Delta t\,, \ee compare
Fig.\ref{curve1}.  In the limit where $\Delta t$ becomes
infinitesimally we obtain the line element
\begin{align}
  d \vc(t) & = (dc^1(t), dc^2(t), dc^3(t)) \nonumber \\ & :=
  \left(\frac{\partial c^1(t)}{\partial t}, \frac{\partial
      c^2(t)}{\partial t}, \frac{\partial c^3(t)}{\partial t}\right)
  \, dt \,.
\end{align}
It is characterized by an infinitesimal length and an orientation.

\vspace{0.5cm}

We now construct objects that we can integrate over the curve $\vc$ 
in order to obtain a coordinate invariant scalar. The line element $d\vc$
contains three independent components $dc^i$. If we shift from old
coordinates $x^i$ to new coordinates $y^{j'}=y^{j'}(x^i)$ these components 
transform according to 
\be
dc^{j'}= \frac{\partial y^{j'}}{\partial x^i} dc^i \,.
\ee
Therefore we can form an invariant expression if we introduce objects
$\valpha=\valpha(x^i)$, with three independent components $\alpha_i$, that 
transform in the opposite way,
\be
\alpha_{j'} = \frac{\partial x^i}{\partial y^{j'}} \alpha_i \,.
\ee
This transformation behavior characterizes a vector or, more precisely,
a covariant vector (a 1-form). It follows that the expression
\be
\alpha_i \, dc^i = \alpha_{j'}\, dc^{j'}
\label{einstein}
\ee
yields the same value in each coordinate system. 

Thus, we can now immediately define integration over a curve by the 
expression
\begin{align}
\int \alpha_i \, dc^i &=  \int \alpha_1\, dc^1 + \alpha_2\, dc^2 + \alpha_3\,
dc^3 \nonumber \\
& = \int \left(\alpha_1 \frac{\partial c^1}{\partial t} +
\alpha_2 \frac{\partial c^2}{\partial t} +
\alpha_3 \frac{\partial c^3}{\partial t}\right) dt \,.
\label{lineintegral}
\end{align}
The last line shows how to carry out explicitly the integration
since $\alpha_i$ and $c^i$ are functions of the parameter $t$.

\subsubsection{Integration over a surface and contravariant vector densities 
as surface integrands}

Now we consider a two-dimensional surface $\vh = \vh(t,s)$. 
Within a specific coordinate system~$x^i$, a parametrization of $\vh$  
is of the form
\be
\vh(t,s) = (a^1(t,s), a^2(t,s), a^3(t,s)) 
\ee
with parameters $t$, $s$ and functions $a^i(t,s)$ that define the shape of 
the surface. 

\vspace{0.5cm}
\begin{figure}[htb]
\begin{center}
\includegraphics[width=0.9\linewidth]{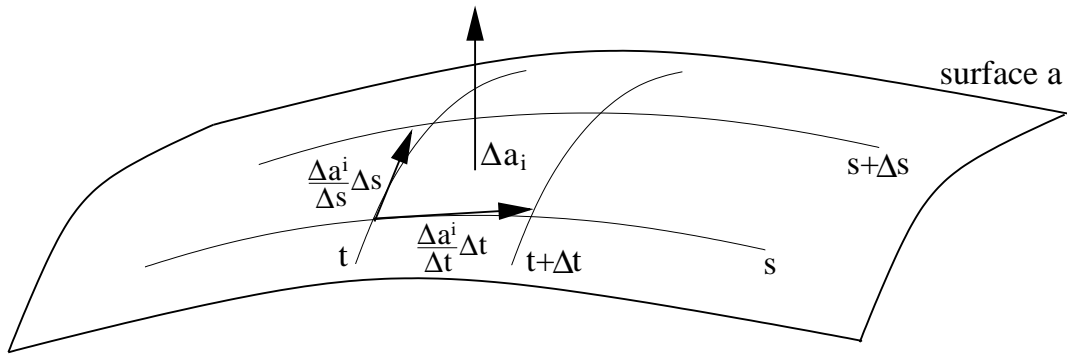}
\caption{Parametrization of a surface $\vh(t,s)$. The lines 
  $t=$const, $t+\Delta t=$const, $s=$const, and $s+\Delta s=$const
  circumscribe a surface $\Delta a_i$ that is spanned by the edges
  $\frac{\Delta a^i}{\Delta t}\, dt$ and $\frac{\Delta a^i}{\Delta
    s}\, ds$.  In the limit $\Delta t \rightarrow dt$, $\Delta s
  \rightarrow ds$, it becomes an elementary surface element $d a_i$.}
\label{surface1}
\end{center}
\end{figure}

An elementary surface element is bound by lines $t=$const,
$t+dt=$const, $s=$const, and $s+ds=$const, compare Fig.\ref{surface1}.
It is characterized by the two edges $\frac{\partial a^i}{\partial
  t}\, dt$ and $\frac{\partial a^i}{\partial s}\, ds$. These edges
span an infinitesimal surface, the area and orientation of which is
characterized by a covariant vector $da_i$ that points normal to the
infinitesimal surface.  The vector $da_i$ is given by the vector
product of $\frac{\partial a^i}{\partial t}\, dt$ and $\frac{\partial
  a^i}{\partial s}\, ds$, \be da_i= \epsilon_{ijk}\frac{\partial
  a^j}{\partial t} \frac{\partial a^k}{\partial s}\, dt\, ds\,.
\label{dhi}
\ee

In order to know how the components $da_i$ transform under coordinate
transformations $y^{j'}=y^{j'}(x^i)$, we have to know the
transformation behavior of the symbol $\epsilon_{ijk}$.  Since in any
coordinate system, $\epsilon_{ijk}$ assumes the values 0, 1, or 
-1 by definition, it is obvious that in general
\be
\epsilon_{i'j'k'} \neq \frac{\partial x^i}{\partial y^{i'}}
\frac{\partial x^j}{\partial y^{j'}}\frac{\partial x^k}{\partial y^{k'}}
\epsilon_{ijk}\,.
\ee
This is because the determinant of the transformation matrix, i.e., 
\be
\det\left(\partial x/\partial y\right) =
\epsilon_{ijk}
\frac{\partial x^i}{\partial y^{i'}}
\frac{\partial x^j}{\partial y^{j'}}\frac{\partial x^k}{\partial y^{k'}}\,,
\label{det}
\ee
is, in  general, not equal to one. But it follows from \eqref{det} that 
the correct transformation rule for $\epsilon_{ijk}$ is given by 
\begin{align}
\epsilon_{i'j'k'} &= \frac{1}{\det(\partial x /\partial y)}
\frac{\partial x^i}{\partial y^{i'}}
\frac{\partial x^j}{\partial y^{j'}}\frac{\partial x^k}{\partial y^{k'}}
\epsilon_{ijk} \nonumber \\
& = \det(\partial y /\partial x)
\frac{\partial x^i}{\partial y^{i'}}
\frac{\partial x^j}{\partial y^{j'}}\frac{\partial x^k}{\partial y^{k'}}
\epsilon_{ijk}\,.
\end{align}
With \eqref{dhi} this yields the transformation rule for the components
$da_i$,
\be
da_{j'} = \det(\partial y /\partial x) \frac{\partial x^i}{\partial y^{j'}}
da_i \,.
\ee

Now we construct quantities that can be integrated over a surface. Since 
a surface element is determined from three independent components
$da_i$ we introduce an integrand with three independent
components $\beta^i$ that transform according to
\be
\beta^{j'} = \frac{1}{\det(\partial y /\partial x)} 
\frac{\partial y^{j'}}{\partial x^i} \beta^i\,.
\label{denstransform}
\ee Transformation rules that involve the determinant of the
transformation matrix characterize so-called densities. Densities are
sensitive towards changes of the scale of elementary volumes. In
physics they represent additive quantities, also called extensities,
that describe how much of a quantity is distributed within a volume or
over the surface of a volume. This is in contrast to intensities. The
covariant vectors that we introduced as natural line integrals are
intensive quantities that represent the strength of a physical field.

The transformation behavior 
\eqref{denstransform} of the components $\beta^i$ characterizes a 
contravariant vector density. With this transformation behavior the 
surface integral
\be 
\int \beta^i da_i = \int \beta^i \epsilon_{ijk}  
\frac{\partial a^j}{\partial t}
\frac{\partial a^k}{\partial s} \, dt\, ds 
\ee
yields a scalar value that is coordinate independent.

\subsubsection{Integration over a volume and scalar densities as 
volume integrands}

We finally consider integration over a three-dimensional volume 
$\vv$ in three-dimensional space. Again we choose a specific
coordinate system $x^i$ and specify a parametrization of
$\vv$ by 
\be
\vv(t,s,r) = \bigl(v^1(t,s,r),v^2(t,s,r),v^3(t,s,r)
\bigr)\,,
\ee
with three parameters $t$, $s$, and $r$.

An elementary volume element $dv$ is characterized by three edges
$\frac{\partial v^i}{\partial t} dt$, 
$\frac{\partial v^i}{\partial s} ds$, and
$\frac{\partial v^i}{\partial r} dr$.
The volume, which is spanned by these edges, is given by the determinant
\begin{align}
dv & = \det\left(\frac{\partial v^i}{\partial t} dt,
\frac{\partial v^i}{\partial s} ds, \frac{\partial v^i}{\partial r} dr
\right) \nonumber \\
& = \epsilon_{ijk}\frac{\partial v^i}{\partial t}
\frac{\partial v^j}{\partial s} \frac{\partial v^k}{\partial r} \,
dt\,ds\,dr
\,.
\end{align}
It is not coordinate invariant but transforms under coordinate transformations
${y^j}'={y^j}'(x^i)$ according to
\be
dv' =  \det(\partial y /\partial x)\, dv\,.
\ee

Since the volume element $dv$ constitutes one independent component, a
natural object to integrate over a volume has one independent
component as well.  We denote such an integrand by $\gamma$. It
transforms according to \be \gamma' = \frac{1}{\det(\partial y
  /\partial x)} \, \gamma \,.  \ee This transformation rule
characterizes a scalar density and yields \be \int \gamma \, dv = \int
\gamma \, \epsilon_{ijk}\frac{\partial v^i}{\partial t} \frac{\partial
  v^j}{\partial s} \frac{\partial v^k}{\partial r} \, dt\,ds\,dr \ee
as a coordinate independent value.

\subsection{Poincar\'e Lemma}

The axiomatic approach takes advantage of the Poincar\'e lemma.  The
Poincar\'e lemma states under which conditions a mathematical object
can be expressed in terms of a derivative, i.e., in terms of a
potential.

We consider integrands $\alpha_i$, $\beta^i$, and $\gamma$ of line-,
surface-, and volume integrals, respectively, and assume that they
are defined in an open and simply connected region of 
three-dimensional space. Then the Poincar\'e lemma yields the following 
conclusions: 
\begin{enumerate}
\item If $\alpha_i$ is curl free, it can be written as the gradient
of a scalar function $f$,
\be
\epsilon^{ijk}\partial_j\alpha_k = 0 \qquad 
\Longrightarrow \qquad \alpha_i =  \partial_i f \,.
\label{poin1}
\ee
\item If $\beta^i$ is divergence free, it can be written as the
curl of the integrand $\alpha_i$ of a line integral,
\be
\partial_i \beta^i  = 0 \qquad \Longrightarrow 
\qquad \beta^i = \epsilon^{ijk} \partial_j \alpha_k \,.
\label{poin2}
\ee
\item The integrand $\gamma$ of a volume integral can be written as the 
divergence of an integrand $\beta^i$ of a surface integral,
\be
\gamma \;\, \text{is a volume integrand} \quad \Longrightarrow 
\qquad \gamma = \partial_i \beta^i \,.
\label{poin3}
\ee
\end{enumerate} 
While conclusions \eqref{poin1}, \eqref{poin2} are familiar from
elementary vector calculus, this might not be the case for conclusion 
\eqref{poin3}. However, \eqref{poin3} is rather trivial since, 
in cartesian coordinates $x$, $y$, $z$, for a given 
volume integrand $\gamma=\gamma(x,y,z)$ the 
vector $\beta^i$ with components $\beta^x=\int_0^{x} \gamma(t,y,z)/3 \, dt$, 
$\beta^y=\int_0^{y} \gamma(x,t,z)/3 \, dt$, and $\beta^z=\int_0^{z} 
\gamma(x,y,t)/3 \, dt$ 
fulfills \eqref{poin3}. Of course, the vector $\beta^i$ is not uniquely
determined from $\gamma$ since any divergence free vector field can be added
to $\beta^i$ without changing $\gamma$. We further note that $\gamma$, as 
a volume integrand, constitutes a scalar density. It can be integrated
as above to yield the components of $\beta^i$ as components of a 
contravariant vector density. Therefore the integration does not yield a 
coordinate invariant scalar such that $\gamma$ cannot be considered
as a natural integrand of a line integral.

\subsection{Stokes Theorem}

In our notation Stokes theorem, if applied to line integrands $\alpha_i$ or
surface integrands $\beta^i$, yields the identities:
\begin{align}
\int_V \partial_i \beta^i \, dv & = 
\int_{\partial V} \beta^i\, da_i \,, \label{stokes1}\\
\int_S \epsilon^{ijk}\partial_j \alpha_k\, da_i & =
\int_{\partial S} \alpha_i \, dc^i \,, \label{stokes2}
\end{align}
where $\partial V$ denotes the two-dimensional boundary of a simply connected
volume $V$ and $\partial S$ denotes the one-dimensional boundary
of a simply connected surface $S$.

\centerline{=========}

\end{document}